\documentclass[final,nonatbib]{article}
\usepackage[numbers]{natbib}
\usepackage{neurips_2024}
\usepackage{macros_aas}
\usepackage[utf8]{inputenc} 
\usepackage[T1]{fontenc}    
\usepackage{hyperref}       
\usepackage{url}            
\usepackage{booktabs}       
\usepackage{amsfonts}       
\usepackage{nicefrac}       
\usepackage{microtype}      
\usepackage[dvipsnames]{xcolor}         
\usepackage{microtype}
\usepackage{graphicx}
\usepackage{subfigure}
\usepackage{booktabs} 
\hypersetup{
	colorlinks   = true, 
    urlcolor     = blue, 
    linkcolor    = blue, 
    citecolor    = red   
 }
\usepackage{amsmath}
\usepackage{amssymb}
\usepackage{mathtools}
\usepackage{amsthm}
\usepackage[capitalize,noabbrev]{cleveref}
\usepackage{algorithm}
\usepackage{algorithmic}
\usepackage{typed-checklist}

\usepackage{xargs}    
\usepackage[textsize=tiny,colorinlistoftodos]{todonotes}

\newcommand{\removethisforspace}[1]{{\color{Purple} 
    \ifthenelse{\equal{\togglemytext}{1}}{#1}{} 
}}

\newcommand{\torch}{{\tt Torch}}
\newcommand{\torchvision}{{\tt Torchvision}}
\newcommand{\numpy}{{\tt Numpy}}

\newcommand{\pytorch}{{\tt PyTorch}}
\newcommand{\matplotlib}{{\tt Matplotlib}}
\newcommand{\sbicodename}{{\tt sbi}\textsubscript{Macke}}
\newcommand{\deeplenstronomy}{{\tt deeplenstronomy}}
\newcommand{\lenstronomy}{{\tt lenstronomy}}

\newcommand{\getdist}{{\tt Getdist}}

\newcommand{\astropy}{{\tt Astropy}}
\newcommand{\hfivepy}{{\tt H5py}}

\newcommand{\python}{{\tt Python}}

\newcommand{\eaf}{{\tt Elastic Analysis Facility (EAF)}}

\newcommand{\myfig}{Fig.}
\newcommand{\mytable}{Table}
\newcommand{\myappendix}{Appendix}
\newcommand{\mysection}{\S}
\newcommand{\myrowspace}{{\color{White}-}}

\newcommand{\modelnpe}{NPE-only}
\newcommand{\modelnpeuda}{NPE-UDA}
\newcommand{\datasource}{source}
\newcommand{\datatarget}{target}
\newcommand{\sersic}{S\'ersic}

\newcommand{\einsteinrad}{\theta_{\mathrm{E}}}
\newcommand{\eccentricity}[2]{e_{\mathrm{#1},#2}}

\newcommand{\myarcsec}{$^{\prime\prime}$}
\newcommand{\myarcsecmath}{^{\prime\prime}}

\newcommand{\lossuda}{L_{\mathrm{UDA}}}
\newcommand{\lossnpe}{L_{\mathrm{NPE}}}
\newcommand{\losstot}{L_{\mathrm{Tot}}}
\newcommand{\lossfactoruda}{\beta_{\mathrm{UDA}}}

\theoremstyle{plain}

\theoremstyle{definition}

\theoremstyle{remark}

\newcommand{\authoraffil}[2]{$^{#1}$ \\ #2} 
\newcommand{\assignaffilnumber}[2]{$^#1$#2} 

\newcommand{\swierc}{Paxson Swierc}
\newcommand{\nord}{Brian D. Nord}
\newcommand{\ciprijanovic}{Aleksandra \'{C}iprijanovi\'{c}}

\newcommand{\tamargo}{Marcos Tamargo-Arizmendi}

\newcommand{\swiercemail}{\url{pswierc@uchicago.edu}}
\newcommand{\nordemail}{\url{nord@fnal.gov}}
\newcommand{\ciprijanovicemail}{\url{aleksand@fnal.gov}}

\newcommand{\tamargoemail}{\url{tamargomarcos@outlook.com}}

\newcommand{\uchicagoAA}{Department of Astronomy and Astrophysics, University of Chicago, Chicago, IL 60637}
\newcommand{\kicp}{Kavli Institute for Cosmological Physics, University of Chicago, Chicago, IL 60637}
\newcommand{\fermilab}{Fermi National Accelerator Laboratory, Batavia, IL 60510}

\newcommand{\ackfundingdeepskies}{We acknowledge the Deep Skies Lab as a community of multi-domain experts and collaborators who’ve facilitated an environment of open discussion, idea generation, and collaboration. This community was important for the development of this project.\\ \\ }

\newcommand{\ackfundingdoegeneral}{Work supported by the Fermi National Accelerator Laboratory, managed and operated by Fermi Research Alliance, LLC under Contract No. DE-AC02-07CH11359 with the U.S. Department of Energy. The U.S. Government retains and the publisher, by accepting the article for publication, acknowledges that the U.S. Government retains a non-exclusive, paid-up, irrevocable, world-wide license to publish or reproduce the published form of this manuscript, or allow others to do so, for U.S. Government purposes.\\ \\ }

\newcommand{\ackfundingdoeecanord}{This material is based upon work supported by the Department of Energy under grant No. FNAL 21-25.\\ \\ }

\newcommand{\contribciprijanovic}{
\'{C}iprijanovi\'{c}: Conceptualization, Methodology, Formal analysis, Writing - Review \& Editing, Supervision, Project administration
\\ \\}

\newcommand{\contribnord}{
Nord: Conceptualization, Methodology, Formal analysis, Resources, Writing - Original Draft, Writing - Review \& Editing, Supervision, Project administration, Funding acquisition
\\ \\}

\newcommand{\contribswierc}{
Swierc: Conceptualization, Methodology, Formal analysis, Software, Validation, Investigation, Data Curation, Writing - Original Draft
\\ \\}

\newcommand{\contribtamargo}{
Tamargo-Arizmendi: Methodology, Formal analysis, Software, Validation, Investigation, Writing - Review \& Editing
\\ \\}

\author{
\swierc\authoraffil{1}{\swiercemail} \And
\tamargo\authoraffil{2}{\tamargoemail} \And
\ciprijanovic\authoraffil{1,2}{\ciprijanovicemail} \And 
\nord\authoraffil{1,2,3}{\nordemail} \\ \\
\assignaffilnumber{1}{\uchicagoAA}\\
\assignaffilnumber{2}{\fermilab}\\
\assignaffilnumber{3}{\kicp}
}

\newcommand{\figurecollage}[3]{
\begin{figure}[htp!]
  \centering
  \includegraphics[scale=1.0]
  {#3}
  \vspace{-0.39cm}
  \caption{#1} #2
\end{figure}
}

\newcommand{\figurefeaturespace}[3]{
\begin{figure}[htp!]
  \centering
  \includegraphics[scale=1.0]
  {#3}
  \vspace{-0.39cm}
  \caption{#1} #2
\end{figure}
}

\newcommand{\tablesetupresults}[2]
{
{\renewcommand{\arraystretch}{1.05}
\begin{table}
   \centering
   \noindent\begin{minipage}[b]{0.99\columnwidth}
   \centering
    \caption{#1} #2
  
  \centering
  \fontsize{9}{9}\selectfont
  \setlength{\tabcolsep}{2.7pt}
  \begin{tabular}{ l c c c c c c }
  \toprule
                            & \multicolumn{2}{c}{\textbf{(a)} Prior Distributions}  & \multicolumn{2}{c}{\textbf{(b)} Residual: Target}                     & \multicolumn{2}{c}{\textbf{(c)} Residual: Source}                                  \\
  \hline   
  Params                    & Training                  & Valid/Test                & \modelnpe                     & \modelnpeuda                  & \modelnpe                     & \modelnpeuda                  \\ 
  \toprule
  $\einsteinrad$ (\myarcsec)& $\mathcal{U}(0.9,3.2)$    & $\mathcal{U}(1.0,3.0)$    & $-0.05 ^{+0.18}_{-0.33}$      & $-0.006 ^{+0.045}_{-0.039}$   & $\myrowspace0.002 \pm 0.017$  & $\myrowspace0.000 \pm 0.018$  \\ 
  $x_\mathrm{}$ (\myarcsec) & $\mathcal{U}(-1.3,1.3)$   & $\mathcal{U}(-1.0,1.0)$   & $-0.06 \pm 0.19$              & $\myrowspace0.007 \pm 0.076$  & $\myrowspace0.004 \pm 0.025$  & $-0.002 \pm 0.030$            \\ 
  $y_\mathrm{}$ (\myarcsec) & $\mathcal{U}(-1.3,1.3)$   & $\mathcal{U}(-1.0,1.0)$   & $\myrowspace0.07 \pm 0.20$    & $\myrowspace0.001 \pm 0.079$  & $-0.001 ^{+0.021}_{-0.024}$   & $-0.001 \pm 0.030$            \\ 
  $\eccentricity{l}{1}$     & $\mathcal{U}(-0.3,0.3)$   & $\mathcal{U}(-0.2,0.2)$   & $-0.28 \pm 0.75$              & $0.007 ^{+0.062}_{-0.075}$    & $-0.001 \pm 0.030$            & $0.001 ^{+0.019}_{-0.016}$    \\
  $\eccentricity{l}{2}$     & $\mathcal{U}(-0.3,0.3)$   & $\mathcal{U}(-0.2,0.2)$   & $\myrowspace0.23 \pm 0.62$    & $0.016 ^{+0.064}_{-0.084}$    & $0.000 ^{+0.015}_{-0.016}$    & $0.004 ^{+0.015}_{-0.020}$    \\ 
  \toprule
  
\end{tabular}
\end{minipage}
\end{table}
}
}

\newcommand{\tableembeddingnetnpe}[2]{
\begin{table}
  \centering
  \noindent\begin{minipage}[b]{\columnwidth}
  \centering
    \caption{#1}
   #2
  \centering
  \begin{tabular}{l l c}
 \hline   Layer   &  Output shape   &  Parameters \\\hline \hline
  Conv2d    &  [-1, 8, 32, 32]  &  $k=3$, $s=1$\\ 
  \midrule
  BatchNorm2d & [-1, 8, 32, 32] & $k=3$, $s=1$\\ 
  \midrule
  Conv2d & [-1, 16, 32, 32] & $k=3$, $s=1$ \\
  \midrule
  BatchNorm2d & [-1, 16, 32, 32] & $k=3$, $s=1$\\ 
  \midrule
  MaxPool2d & [-1, 16, 16, 16] & $k=2$, $s=2$  \\
  \midrule
  Conv2d    &  [-1, 32, 16, 16]  &  $k=3$, $s=1$\\ 
  \midrule
  BatchNorm2d & [-1, 32, 16, 16] & $k=3$, $s=1$\\ 
  \midrule
  Conv2d & [-1, 32, 16, 16] & $k=3$, $s=1$\\
  \midrule
  BatchNorm2d & [-1, 32, 16, 16] & $k=3$, $s=1$\\ 
  \midrule
  MaxPool2d & [-1, 32, 8, 8] & $k=2$, $s=2$ \\
  \midrule
  Conv2d    &  [-1, 64, 8, 8]  &  $k=3$, $s=1$ \\ 
  \midrule
  BatchNorm2d & [-1, 64, 8, 8] & $k=3$, $s=1$\\ 
  \midrule
  Conv2d & [-1, 128, 8, 8] & $k=3$, $s=1$\\
  \midrule
  BatchNorm2d & [-1, 128, 8, 8] & $k=3$, $s=1$\\ 
  \midrule
  MaxPool2d & [-1, 128, 4, 4] & $k=2$, $s=2$ \\
  \midrule
  Flatten & [-1, 2048] & - \\
  \midrule
  Linear & [-1, 20] & - \\
  \hline
\end{tabular}
\end{minipage}
\end{table}
}

\title{
Domain-Adaptive Neural Posterior Estimation for Strong Gravitational Lens Analysis
}

\begin{document}
\maketitle


\begin{abstract}
Modeling strong gravitational lenses is prohibitively expensive for modern and next-generation cosmic survey data.
Neural posterior estimation (NPE), a simulation-based inference (SBI) approach, has been studied as an avenue for efficient analysis of strong lensing data. 
However, NPE has not been demonstrated to perform well on out-of-domain target data --- e.g., when trained on simulated data and then applied to real, observational data. 
In this work, we perform the first study of the efficacy of NPE in combination with unsupervised domain adaptation (UDA). 
The \datasource{} domain is noiseless, and the \datatarget{} domain has noise mimicking modern cosmology surveys.
We find that combining UDA and NPE improves the accuracy of the inference by 1-2 orders of magnitude and significantly improves the posterior coverage over an NPE model without UDA.
We anticipate that this combination of approaches will help enable future applications of NPE models to real observational data.
\end{abstract}

\section{Introduction and Related Work}
\label{sec:introduction}
Galaxy-scale strong gravitational lensing is a cosmic probe that provides key information about dark energy, dark matter, and galaxy evolution~\cite{albrecht_report_2006, 2023arXiv230705714T, 2010ARA&A..48...87T, 2004PhRvD..70d3534L, 2023MNRAS.524.6159K, 2024MNRAS.tmp.1779G, 2024ApJ...970..143T, 2015salt.confE..16S}.
Modern and future cosmic survey experiments --- e.g., the Dark Energy Survey (DES)~\cite{the_dark_energy_survey_collaboration_dark_2005, flaugher_dark_2015},  Hyper Suprime-Cam~\cite{2022PASJ...74..247A, 2023PhRvD.108l3520M}, the Kilo-Degree Survey (KiDS)~\cite{2013ExA....35...25D, 2019A&A...625A...2K}, the Rubin Observatory Legacy Survey of Space and Time~\cite{ivezic_lsst_2019}, Euclid~\cite{2024arXiv240513491E}, JWST~\cite{schaerer_first_2022, brummel-smith_inferred_2023}, and the Nancy Grace Roman Telescope~\cite{eifler_cosmology_2021, la_plante_prospects_2023, wang_high_2022-1} --- are expected to contain 10$^3$ - 10$^5$ lensing systems~\cite{2016ApJ...827...51N, 2024arXiv240608919S, Collett2015}.
Traditional techniques for lens modeling have relied heavily on analytic likelihood-fitting, which is computationally expensive and human-time intensive~\cite{Lefor_2013}. 
Additionally, due to simplifying assumptions in designing the likelihoods, these techniques often lack the capability of modeling non-Gaussian likelihoods and posteriors~\cite{Lefor_2013}. 
However, these techniques have advanced notably in automation and speed~\cite{2021JOSS....6.2825N, 2022ApJ...935...49G, 2021A&A...647A.176G, 2021MNRAS.503.2380S}.
Supervised deep learning-based inference techniques --- including neural network regression and the recently reinvigorated simulation-based inference (SBI)~\cite{cranmer_frontier_2020, groenke_bgroenks96simulationbasedinferencejl_2024, noauthor_probabilistslampe_2024, dirmeier_dirmeiersbijax_2024, zhang_nbi_2023, gloeckler_all--one_2024, haggstrom_fast_2024} like neural posterior estimation (NPE)~\cite{papamakarios_masked_2018, greenberg_automatic_2019, wang_preconditioned_2024, zeghal_neural_2022-1} --- have been studied in applications on a wide variety of physics and cosmology topics~\cite{danilov_neural_2022, khullar_digs_2022, poh_strong_2022, reza_estimating_2022, barret_simulation-based_2024-1, hahn_accelerated_2022-1, 2024PhRvD.109f4056L}, including strong lensing~\cite{2024arXiv240717292J}.
Once these models are trained (aka, ``amortized''), these methods are very fast compared to traditional modeling methods~\cite{cranmer_frontier_2020}.
In many areas of cosmology, including strong lensing, when there isn't enough real observational data for training deep learning-based models, realistic simulations are used~\cite{2015ascl.soft05026T, deeplenstronomy, lenstronomy:2018, lenstronomy:2021}.
Nevertheless, these simulated data can differ significantly from real, observational data --- i.e., observational noise, astrophysics, and cosmology.
The differences between the simulated training data (\datasource{} domain) and the real observational data used for analysis (\datatarget{} domain) constitute domain shifts between data distributions that can cause models to favor the \datasource{} domain~\cite{2019arXiv190911575S, 2024arXiv240408184V, 2023arXiv230519499M}. Studies of model misspecification due to domain shift ~\cite{Ward_2022, Wang_2023, Canon_20024, Schmitt_2024} have shown this to be a significant limitation of SBI and its application to out-of-domain data.
Domain adaptation (DA) is a class of deep learning techniques that help neural networks adapt to domain shifts so that the feature spaces of the \datasource{} and \datatarget{} domains align during training~\cite{xu_video_2022, triess_survey_2021, zhuang_comprehensive_2020, zhang_transfer_2020, kouw_introduction_2019-1, le_deep_2019}.
Unsupervised domain adaptation (UDA) does not require labels on the \datatarget{} data~\cite{wilson_survey_2020, kouw_review_2021, schwonberg_survey_2023, zhao_review_2020}.
This has been studied as an approach to ameliorate biases due to domain shifts for neural network-based analyses in many problems, including cosmology and strong lensing~\cite{cabrera-vives_domain_2023, ciprijanovic_domain_2020, dellaiera_deep_2023, ciprijanovic_deepmerge_2021, swierc_domain_2023, kim_deep_2021, farrens_deep_2022, ciprijanovicRobustnessDeepLearning2021, roncoli_domain_2024, 10.1093/mnras/sty3497, astronomy3030012, 2021A&A...653A..22K}.    
In this work, we advance the state of the art by combining NPE and UDA and comparing the performance of \modelnpeuda{} and \modelnpe{} models on strong lensing data in two different domains, which are distinguished by the noise in the images.

\section{Methods: Lensing, Neural Posterior Estimation,  Domain Adaptation}
\label{sec:methods}
\textbf{Physics of strong gravitational lensing:} 
When light from a background object encounters a sufficiently massive lensing object on its way to an observer, the image of the background object is significantly magnified and distorted ~\cite{1996astro.ph..6001N, 1998LRR.....1...12W}.
This warped image is the primary observable data (see \myfig{}~\ref{fig:collage}(b) for example images). 
In parametric lens modeling, one can consider $>10$ parameters from the background object and the lens that could be inferred from the imaging data~\cite{Lefor_2013, Keeton_2016, 10.1093/mnras/stab1547, Birrer_2015}. 
In this study, we infer only five parameters related to the lens: Einstein radius $\einsteinrad$, relative angular positions between the background object and lens ($x$, $y$), and lens eccentricity moduli ($\eccentricity{l}{1}$,$\eccentricity{l}{2}$). 
Like all astronomical data, strong lensing images are subject to observational noise from multiple sources --- e.g., atmosphere, sky brightness, CCD gain, number of exposures, exposure time, CCD readout, and photon counting.
These noises can add values to pixel counts or cause blurring in the images; they need to be accounted for in model building to avoid systematic bias and large error bars.

\textbf{Neural Posterior Estimation (NPE):} 
To infer parameter posterior densities, we employ NPE~\cite{greenberg_automatic_2019}, which uses a CNN-based embedding network to summarize images into features, which are then passed to a Masked Autoregressive Flow (MAF), a combination of an autoregressive model and a normalizing flow~\cite{papamakarios_masked_2018}, to estimate posterior densities. 
MAF can estimate posterior distributions of arbitrary shape (i.e., non-Gaussian).
In the standard \modelnpe{} approach, there is a single loss function $\lossnpe{}$ that takes the form of the negative log posterior volume~\cite{greenberg_automatic_2019}.

\textbf{Unsupervised Domain Adaptation (UDA):} 
In UDA methods, the source domain data have labels, and the target domain data do not have labels.
Common UDA approaches include adversarial methods~\cite{swierc_domain_2023, mathelin_adversarial_2021, guAdversarialReweightingPartial1969, kim_instance-based_1969, jiang_unsupervised_2023, ganin_domain-adversarial_2016} and distance-based methods ~\cite{wilson_survey_2020, farahani_brief_2020}. 
In distance-based methods, the loss is defined as a multi-dimensional distance between latent features from the source and target domain data.
In this work, we use distance-based methods, for which the UDA loss function $\lossuda{}$ is the Maximum Mean Discrepancy (MMD)~\cite{gretton2012kernel}.
MMD is a method that calculates the distance between distributions: when applied to the latent feature space, it can be used as a loss function. In~\cite{Schmitt_2024},  MMD was used as a metric to quantify the NPE model misspecification.
When MMD is included as a loss during training, it is intended to cause the network to align latent feature spaces for the source and target data; this leads to the extraction of domain-invariant features and enables the model to work well on both.

\textbf{Combining NPE and UDA:} 
We combine NPE and UDA methods via their losses.
The UDA loss $\lossuda{}$ is calculated using the source and target domain latent features (without labels) at the end of the embedding network. 
The NPE loss $\lossnpe{}$ is calculated using the \datasource{} data (with labels) at the end of the MAF. 
The total loss function $\losstot{} = \lossnpe{} + \lossfactoruda{} * \lossuda{}$ is used with gradient descent to update all weights; $\lossfactoruda{}$ is a hyperparameter weighting the MMD loss.

\section{Experiments}
\label{sec:experiments}

\textbf{Data:}
We use the \deeplenstronomy~\cite{deeplenstronomy} software, which is built on \lenstronomy~\cite{lenstronomy:2018, lenstronomy:2021}, to generate simulations of galaxy-scale strong lensing images as if observed in a ground-based survey.
We use a single photometric band ($g$), which is sufficient for producing morphological features of a lensing system.
Images have a pixel scale of $0.263$ \myarcsec{}/pixel to match that of DES~\cite{flaugher_dark_2015, DESdatarelease1}.
During simulation, the surface brightness of the lensing galaxy is omitted from the images: this exclusion represents a part of the typical lens modeling process in which lens light is removed before the lensed background image is modeled~\cite{levasseur_uncertainties_2017}.
We use empirically and theoretically motivated uniform priors for distributions of physics parameters of the background object and the lens object.
For the background object parameters, which we don't infer, we use the following: \sersic{} index $n\sim\mathcal{U}(2,4)$, scale radius $R\sim\mathcal{U}(0.5\myarcsecmath{},1\myarcsecmath{})$, two-dimensional eccentricity $\{\eccentricity{s}{1}, \eccentricity{s}{2} \}\sim\mathcal{U}(-0.2,0.2)$; two-dimensional external shear $\{\gamma_{1}, \gamma_{2}\}\sim\mathcal{U}(-0.05,0.05)$ ~\cite{Collett2015, SLACS4}. 
The apparent magnitude of the background object has a distribution $m \sim\mathcal{U}(22.5,23)$, which is faint enough that the noise will be apparent.
For the lens object parameters that we infer ($\einsteinrad$, $x$, $y$, $\eccentricity{l}{1}$, and $\eccentricity{l}{2}$), the prior distributions are shown in \mytable{}~\ref{table:setupresults}.

We incur a shift in the domain between the source and the target in terms of image noise characteristics only --- i.e., not for the physics parameters. 
The \datasource{} data has noise characteristics that represent a relatively noiseless image: read noise is zero e$^-$, CCD gain is 6.083 e$^-$/count, exposure time is 90 seconds (typical modern optical cosmic surveys), number of exposures is 10, magnitude zero point is 30, sky brightness is 23.5 magnitude/arcsec$^2$ (dimmer than the source light profile), and seeing is 0.9\myarcsec{} (moderate for modern optical cosmic surveys).
In contrast, the \datatarget{} data has noise characteristics that mimic those of the DES: read noise is 7.0 e$^-$, and exposure time, number of exposures, magnitude zero point, sky brightness, and seeing are sampled from empirical distributions~\cite{DESdatarelease1}; these distributions are encoded in the \deeplenstronomy{} package. 

The training set contains 200,000 images in each domain --- source and target. 
These are drawn from the training priors.
The validation and test sets each contain 1,000 images in each of the domains. 
These are drawn from the test priors.
The \sbicodename{} package holds out 10\% of the training data for validation during training and early stopping; that validation set is independent of the one we created. 
For the lens parameters that we infer, the prior distributions for training are wider than the prior distributions for testing to mitigate biases near the edges of the test distribution (see \mytable{}~\ref{table:setupresults}).
The test set is used for all results and metrics in this paper.
Sample lenses from source and target data are shown in \myfig{}~\ref{fig:collage}(b).
All images are 32$\times$32 pixels in shape.
The data set uses $\sim3.5$ GB of storage space.
The data used in this project can be provided upon request.

\textbf{Model Optimization:}
We use the \sbicodename{} package~\cite{tejero-cantero_sbi_2020}, which utilizes PyTorch~\cite{pytorch} to perform NPE analyses.
For the NPE model, we use an embedding network to summarize the image data before input to the MAF. 
The embedding network architecture has six convolution blocks (each with a convolution, max-pooling, and batch normalization layer) followed by one dropout (rate is 0.5) and one dense layer with 20 nodes. 
The MAF has 20 transformation blocks, with 400 hidden features in each. 
This NPE architecture was introduced in~\cite{poh_2024_sbi}.
We experimented with a variety of hyperparameter choices and data sets. 
We determined that the \sbicodename{} package defaults most clearly show the models' performances: the batch size, learning rate, optimizer, and early stopping epochs are 50, 0.0005, \texttt{Adam} optimizer ~\cite{kingma2017adam}, and 20, respectively. 
We set $\lossfactoruda{} = 1.0$. 
We discuss computational costs for model training in \myappendix~\ref{sec:app:computationcosts}.
The code for this work can be provided upon request.%

\section{Results: UDA improves NPE performance on target domain data}
\label{sec:results}

\tablesetupresults{
Distributions and results for each lensing parameter.
The parameters (``params''); \textbf{(a)} prior distributions for training and test sets; \textbf{(b)}  the residuals for the \modelnpe{} and the \modelnpeuda{} models applied to the target domain data; \textbf{(c)} the mean residuals for the \modelnpeuda{} model applied to the source domain data.
}{\label{table:setupresults}}

First, we check that the addition of UDA to NPE does not lead to a significant deterioration in model performance on source data compared to NPE alone.
For all parameters, the \modelnpeuda{} model is slightly more accurate when applied to the \datasource{} data than when applied to the \datatarget{} data (\myfig{}~\ref{fig:collage}(a) and \mytable{}~\ref{table:setupresults}(c)). 
Also, the \modelnpeuda{} model has nearly the same degree of calibration on \datasource{} data as on \datatarget{} data (\myfig{}~\ref{fig:collage}(d)).
Next, the demonstration of performance relies on comparing the \modelnpe{} and \modelnpeuda{} models on \datatarget{} data.
The \modelnpe{} model has an average residual (i.e., bias) of $0.26$\myarcsec{}\ for the Einstein radius.
This bias is far outside the state-of-the-art uncertainties for traditional modeling techniques, which produce uncertainties at the level of $\sim0.01$\myarcsec{} ~\cite{10.1093/mnras/stad3514} or, more generally, at $\sim 1$-$5\%$ ~\cite{2011ApJ...727...96R, 2013ApJ...777...97S}.
In contrast, for the \modelnpeuda{} model, the accuracy improves (the average residual reduces) by approximately 88\%, 88\%, 99\%, 98\%, and 93\% for all five parameters $\einsteinrad$, $\eccentricity{l}{1}$, $\eccentricity{l}{2}$, $x$, $y$, respectively. 
Also, in applications to the target domain data, the parameter uncertainties for the \modelnpeuda{} model are very well-calibrated, while those for the \modelnpe{} model are highly overconfident (\myfig{}~\ref{fig:collage}(d)).
This reflects the \modelnpe{} model's bias toward the source domain when applied to the target domain data.
Finally, the feature spaces for the \modelnpeuda{} model applied to \datasource{} and \datatarget{} data are overlapping but not when the \modelnpe{} model is applied to data from those domains (\myfig{}~\ref{fig:collage}(b) left and right, respectively).

\figurecollage{
  \textbf{(a)}: Two example images from the source (without noise; left) and target (with noise; right) domains.
  \textbf{(b)}: Feature spaces of the embedding network when models are applied to the source (points are filled circles; all points encompassed by a blue circle) and target (points are filled triangles; all points encompassed by an orange circle) domain data for the \modelnpe{} (left) and \modelnpeuda{} (right) models, respectively. 
  \textbf{(c)}: Residuals on the five lens parameters ($\einsteinrad$, $x$, $y$, $\eccentricity{l}{1}$, $\eccentricity{l}{2}$) for the \modelnpe{} model applied to \datatarget{} data (orange), the \modelnpeuda{} model applied to \datatarget{} data (blue), and the \modelnpeuda{} model applied to \datasource{} data (pink). 
  Contours show the 68th- and 95th-percentile confidence regions, and the dashed lines show zero residuals.
  \textbf{(d)}: Posterior coverage on the five lens parameters for the \modelnpe{} model for the \modelnpeuda{} model applied to \datatarget{} data (dashed, color), the \modelnpe{} model applied to \datatarget{} data (solid, color), and the \modelnpeuda{} model applied to \datasource{} data (solid, black). 
  The boundary between underconfident (upper) and overconfident (lower) is marked by a dotted gray line.
}{\label{fig:collage}}{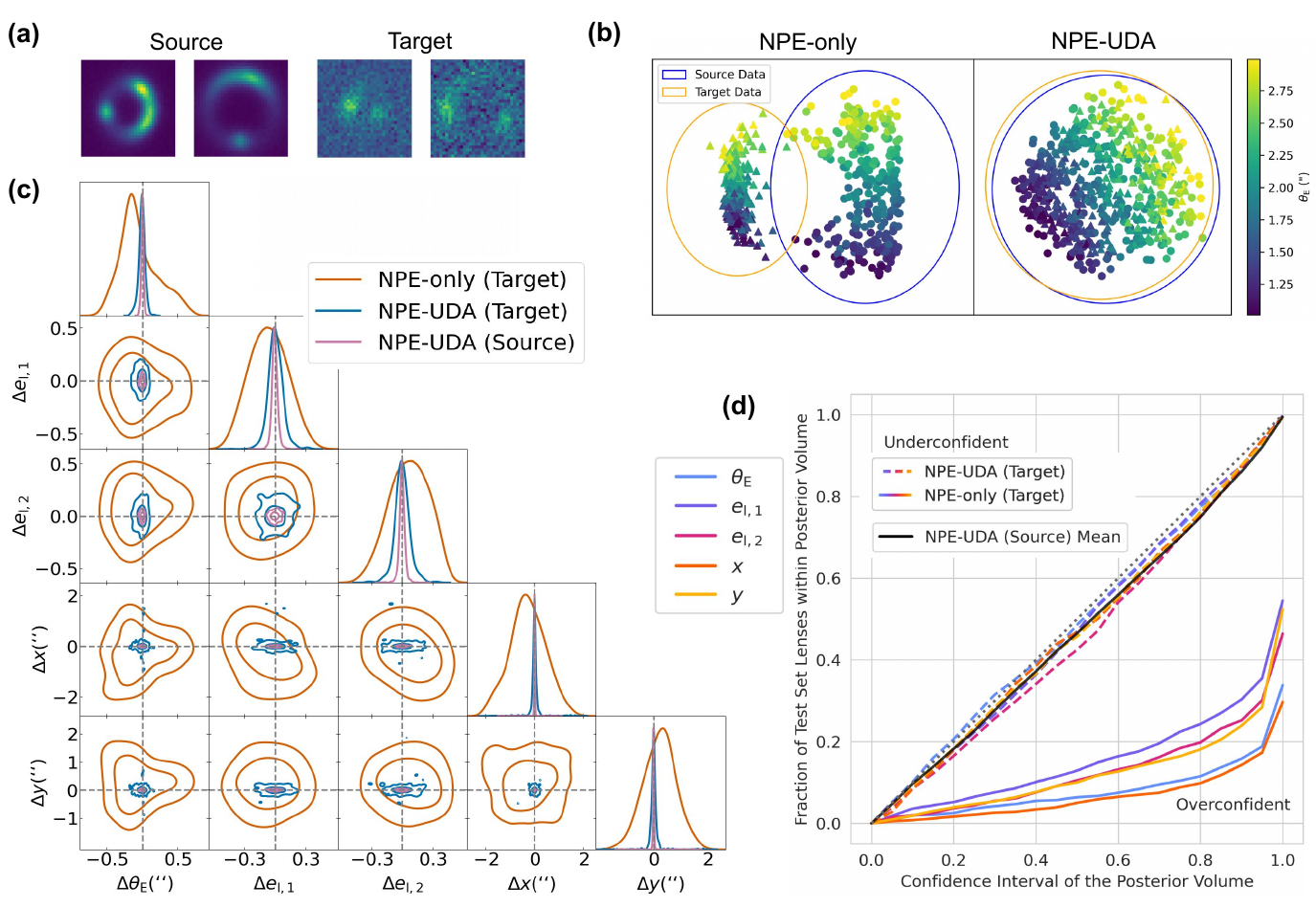}

\section{Summary and Outlook}
\label{sec:conclusion}

We show for the first time that (unsupervised) domain adaptation (UDA)  enhances simulation-based inference (SBI) models when applied to unlabeled \datatarget{} domain data.
We used neural posterior estimation (NPE) to infer five parameters of lensing systems from single-band imaging data.
We compare NPE models that have UDA (\modelnpeuda{}) to NPE models that don't have UDA (\modelnpe{}) (\mysection{}\ref{sec:methods}).
We incurred a domain shift between the \datasource{} and \datatarget{} domains: the \datasource{} images are nearly noiseless, and the \datatarget{} images have the same noise characteristics as DES (\mysection{}\ref{sec:experiments}). 
When applied to the \datatarget{} domain, the \modelnpeuda{} model is 1-2 orders of magnitude more accurate than the \modelnpe{} model for all five lens parameters (\myfig{}~\ref{fig:collage}(c) and \mytable{}~\ref{table:setupresults}(b)). 
Similar approaches may significantly improve the accuracy of SBI/NPE models when they are applied to real observational data.

\bibliography{neurips_custom_2024, neurips_general_2024, neurips_software_2024}
\bibliographystyle{plain}

\newpage
\appendix
\onecolumn


\begin{ack} 

\section{Funding}
\label{sec:app:funding}

\ackfundingdeepskies
\ackfundingdoegeneral
\ackfundingdoeecanord

\section{Author Contributions}
\label{sec:app:funding}

\contribswierc
\contribtamargo
\contribciprijanovic
\contribnord

We thank the following colleagues for their insights and discussions during the development of this work: Jason Poh.

\end{ack}

\section{Attributions: Software and Computing Facilities}
\label{sec:app:software}

We used the following software packages:
\astropy{}~\cite{astropy:2013,astropy:2018,astropy:2022},
\deeplenstronomy{}~\cite{deeplenstronomy},
\eaf{}~\cite{eaf},
\getdist{}~\cite{getdist},
\hfivepy{}~\cite{h5py},
\lenstronomy{}~\cite{lenstronomy:2018, lenstronomy:2021},
\matplotlib{}~\cite{matplotlib},
\numpy{}~\cite{numpy},
\python{}~\cite{python},
\pytorch{}~\cite{pytorch},
\sbicodename{}~\cite{tejero-cantero_sbi_2020},
\torch{}~\cite{torch},
\torchvision{}~\cite{torchvision2016}.


\section{Embedding Network}
\label{sec:app:embedding}
\tableembeddingnetnpe{
The architecture of the embedding network used in the NPE to compress the image data into summary features. 
The first column lists the layer type, the second column lists the dimensionality of the output from that layer, and the third column lists the parameters of that layer; $k$ is the kernel size, and $s$ is the stride. 
The final layer outputs the summary features.}
{\label{table:embedding_net}}

We use an embedding network to reduce the feature space of the imaging data before the MAF uses it. 
See \mytable{}~\ref{table:embedding_net} for the architecture setup, including hyperparameters.

\section{Additional Feature Space Inspection}
\label{sec:app:featurespace}

\figurefeaturespace{
Latent space of the embedding network when NPE is applied to the source and target domain test set data for the \modelnpe{} (left) and \modelnpeuda{} (right) models, respectively.   
This is applied to parameters $x$ \textbf{(a)}, $y$ \textbf{(b)}, $\eccentricity{l}{1}$ \textbf{(c)}, $\eccentricity{l}{2}$ \textbf{(d)}.
}
{\label{fig:latentappendix}}
{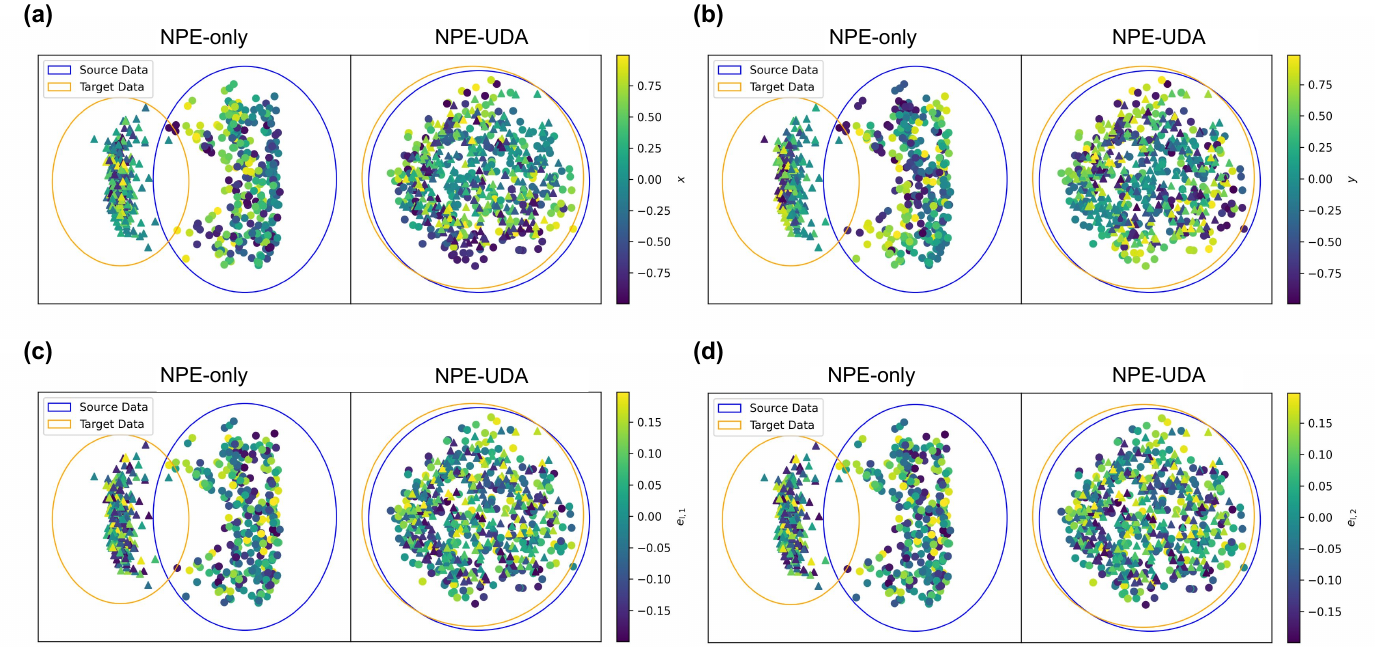}

In the main text, we showed that the feature spaces for the \modelnpeuda{} model on the source and target domains are well-aligned (\mysection{}\ref{sec:results} and \myfig{}~\ref{fig:collage}(b)) for Einstein radius, and there is a clear monotonic correlation between the feature space and the Einstein radius magnitude. 
Here, we further inspect the feature spaces for the \modelnpe{} and \modelnpeuda{} models on the relative positions $x$ and $y$, and on the lens eccentricity moduli $\eccentricity{1}{1}$ and $\eccentricity{l}{2}$.
A non-monotonic correlation exists between the feature space and the parameter of interest for $x$ and $y$, but not for the lens eccentricities. 
We experimented with the isomap hyperparameter for the number of neighbors (default is five), and we found that values greater than 20 did not change the visualization.
We speculate that while the \modelnpeuda{} model requires the feature spaces to overlap, it prioritizes the correlation with the Einstein radius over other parameters. 
Additionally, the other parameters are more subtly represented in the images and thus may be more difficult to learn.


\section{Computational costs for experiments}
\label{sec:app:computationcosts}

All computing was executed on an NVIDIA A100 GPU with 10GB memory. 
These computations were performed on the Fermilab Elastic Analysis Facility~\cite[EAF;][]{eaf}.
Training without UDA requires ~$\sim4.0$ hours, while training with UDA requires ~$\sim6.5$ hours.
This additional time is primarily due to a) calculating the additional loss function for UDA and b) using twice the amount of data by including the target domain data.

\end{document}